\documentclass[12pt,a4paper]{article}

\usepackage{feynmp,amsbsy,latexsym,cite}

\newcommand{\no}{\noindent}

\newcommand{\de}{\delta}
\newcommand{\ep}{\epsilon}

\newcommand{\la}{\lambda}
\newcommand{\si}{\sigma}
\newcommand{\Ga}{\Gamma}

\newcommand{\eps}{\epsilon}

\newcommand{\ket}[1]{\vert #1 \rangle}
\newcommand{\bra}[1]{\langle #1 \vert}

\newcommand{\tr}{\mathrm{tr}\,}
\newcommand{\xb}{{\boldsymbol x}}

\newcommand{\zb}{{\boldsymbol z}}
\newcommand{\wb}{{\boldsymbol w}}

\newcommand{\qb}{{\boldsymbol q}}
\newcommand{\rb}{{\boldsymbol r}}
\newcommand{\kb}{{\boldsymbol k}}
\newcommand{\lb}{{\boldsymbol l}}

\newcommand{\Ab}{{\boldsymbol A}}
\newcommand{\Eb}{{\boldsymbol E}}
\newcommand{\Bb}{{\boldsymbol B}}
\newcommand{\nablab}{{\boldsymbol\nabla}}

\newcommand{\inttwo}[1]{\int \frac{\mathrm{d}^2{#1}}{(2\pi)^2}}

\begin{document}

\begin{titlepage}

\rightline{PLY-MS-00-31}
\rightline{UAB-FT-00-41}

\vskip19truemm
\begin{center}{\Large{\textbf{The Structure of the  QCD Potential in $\mathbf{2+1}$
 Dimensions}}}\\ [12truemm] \textsc{Emili
Bagan}\footnote{email: bagan@ifae.es}\\ [5truemm]
\textit{Dept.~Fisica Te\`orica \&\ IFAE\\
Edifici Cn\\
Universitat Aut\`onoma de Barcelona\\E-08193 Bellaterra  (Barcelona)\\
Spain}\\ [12truemm]\textsc{Martin
Lavelle}\footnote{email: mlavelle@plymouth.ac.uk} and
\textsc{David McMullan}\footnote{email:
dmcmullan@plymouth.ac.uk}\\ [5truemm] \textit{Department of
Mathematics and Statistics\\ The University of Plymouth\\
Plymouth, PL4 8AA\\ UK} \end{center}

\bigskip\bigskip\bigskip
\begin{quote}
\textbf{Abstract:} We calculate the screening and anti-screening
contributions to the inter-quark potential in $2+1$ dimensions, which
is relevant to the high temperature limit of QCD. We  demonstrate
that the relative strength of screening to anti-screening agrees with
the $3+1$ dimensional theory to better than one percent accuracy.
\end{quote}

\end{titlepage}

\setlength{\parskip}{1.5ex plus 0.5ex minus 0.5ex}

\noindent Sparked in part by the discovery of the top quark, there has recently
been a great deal  of interest in the inter-quark
potential, see, for
example,~\cite{Peter:1997ig,Peter:1997me,Lavelle:1998dv,
Schroder:1998vy,Brambilla:1999qa,Brodsky:1999fr,Schroder:1999sg}.
 It has long been
known~\cite{Nielsen:1981sx,Hughes:1980ms,Gribov:1977mi,Drell:1981gu,Vainshtein:1982wh}
that the pure QCD
corrections to the Coulombic potential in 3$+$1 dimensions are of two
types: a dominant anti-screening contribution and a lesser
interaction which corresponds to screening by physical, transverse gluons.
In this letter we will continue our programme~\cite{Lavelle:1997ty,Lavelle:1998dv}
to study the structure of the forces between quarks.
We will demonstrate that in $2+1$ dimensions, which for Euclidean metrics is by dimensional
reduction  related~\cite{Appelquist:1981vg,Nadkarni:1983kb} to the high temperature limit of QCD,
the inter-quark potential has an unexpectedly rich structure and
that the relative weights of the attractive and repulsive
interactions are almost identical to those of the $3+1$ case.

In $SU(N)$, for static quarks without additional light fermions, the
inter-quark potential to order $g^4$ in $d+1$ space-time dimensions
is given by \cite{Schroder:1999sg}
\begin{equation}
V(q )=-\frac{g^2C_F}{\qb^2}\left\{1+g^2\mu^{2\eps}C_A(4d-1)
\frac{|\qb|^{d-3}}{(16\pi)^{\frac{d}{2}}}
\frac{\Gamma(\frac{3-d}{2})\Gamma(\frac{d+1}{2})}{\Gamma(\frac{d}2+1)
}
\right\}\,,
\end{equation}
where $q=|\qb|$, $C_A=N$, $C_F=(N^2-1)/(2N)$ and
$d+1=4-2\eps$. In $3+1$ dimensions this
reduces to the familiar form~\cite{Appelquist:1977tw}
\begin{equation}
V(q)=-\frac{g^2C_F}{\qb^2}\left\{1-{g^2\over(4\pi)^2}C_A\,
{11\over3}\ln\left(\frac{\qb^2}{\mu^2}\right) \right\}\,,
\end{equation}
while in $2+1$ dimensions this becomes~\cite{Schroder3d}
\begin{equation}\label{vtwolow}
V(q)=-\frac{g^2C_F}{\qb^2}\left\{1+g^2C_A
\frac{7}{32|\qb|}\right\}\,.
\end{equation}
Note that in $2+1$ dimensions the result is finite and no
renormalisation is needed \cite{Schroder:1999sg}. In $3+1$ dimensions the order
$g^4$ correction to the Coulombic potential is related to the
universal beta function of QCD, but in $2+1$ dimensions no such
identification is possible since the beta function vanishes.
Finally, we recall that in $2+1$ dimensions the coupling constant
is a dimensionful quantity.

These corrections to
the potential have been  understood in 3$+$1 dimensions as the sum of two
distinct physical effects: a dominant anti-screening interaction
which arises from the Coulombic potential, and a smaller screening
interaction which arises from the virtual production of physical,
i.e., gauge invariant,
gluon pairs. The dominance of anti-screening over screening is the
origin of QCD's asymptotic freedom. Concretely the coefficient of the
logarithmic correction can be decomposed as:
\begin{equation}
V(q)=-\frac{g^2C_F}{\qb^2}\left\{1-{g^2\over(4\pi)^2}C_A\,
\left[4-\frac13\right]\ln\left(\frac{\qb^2}{\mu^2}\right) \right\}\,,
\end{equation}
where the factor of $4$ comes from the anti-screening interaction
and the $\frac13$ from the smaller screening forces. The relative strength
of the screening part of the potential is, we note,  only $8.33\%$ of the
anti-screening contribution.  Due to the universality of the beta
function, this decomposition can be calculated in many different
ways~\cite{Nielsen:1981sx,Hughes:1980ms,Gribov:1977mi,Drell:1981gu,Vainshtein:1982wh,Brown:1997gm},
although it cannot be obtained from the Wilson
loop approach to the potential. This structure of the inter-quark
potential was previously unknown in $2+1$ dimensions and we shall
now calculate it.

We will follow the method of Ref.~\cite{Lavelle:1998dv}. The lowest
energy states corresponding to two heavy quarks a distance $r=|\rb|$ apart
is $\ket{\bar\psi(\rb)h(\rb)h^{-1}(0)\psi(0)}$, where the quarks are
in the same time slice. We call $h^{-1}$ a dressing for the matter
field, $\psi$. This field dependent term  is the lowest energy
gluonic configuration around an individual fermion which maintains
gauge invariance for the composite charged quark. The kinematics of
the heavy quark determines the form of the dressing, and we have
shown elsewhere~\cite{Bagan:1998kg} that it factors into a product
of two terms: a gauge
dependent term which makes the dressed quark gauge invariant, and a
gauge invariant structure.

The first part of the dressing is the minimal gluonic configuration
which renders the quark gauge invariant.
This first term originates from Gauss' law and hence from
longitudinal degrees of freedom. It produces the spreading of the
colour charge, \emph{anti-screening}, which underlies asymptotic freedom in non-abelian
gauge theories, and will raise the energy of the quark-antiquark
state. It is the non-abelian extension of the Coulomb
interaction. Since the overall dressed quark has to correspond to the
lowest energy state, the  additional gauge invariant glue must lower
the energy. As such, it can only correspond to a \emph{screening}
contribution. This physical decomposition into structures which
necessarily raise and lower the energy is the correct identification
of anti-screening and screening effects even in 2$+$1 dimensions
where the coupling does not run.

Generalising the construction in~\cite{Lavelle:1998dv} to $d$-spatial dimensions,
the anti-screening part of the potential at order $g^4$ is
\begin{eqnarray}\label{vd}
V^{(4)}_{\mathrm{anti}}(r)&=&-3g^4C_AC_F\frac{\Ga^3(\frac{d}2-1)}{64
\pi^{\frac{3d}2}}\int\!\mathrm{d}^dz\,\mathrm{d}^dw
\frac1{|\zb-\wb|^{d-2}}\times\nonumber\\
&&\qquad\qquad\qquad\left(\partial^z_j\frac1{|\zb-\rb|^{d-2}}\right)
\left(\partial^w_k\frac1{|\wb|^{d-2}}\right)\bra{0}A^{\mathrm{T}}_j(\zb)A^{\mathrm{T}}_k(\wb)\ket{0}\,,
\end{eqnarray}
where we have used the $d$-dimensional result
\begin{equation}\label{nablad}
\left(\frac1{\nabla^2}f\right)(x^0,\xb)=
-\frac{\Gamma(\frac{d}{2}-1)}{4\pi^{\frac{d}2}}
\int \mathrm{d}^dz\frac{f(x^0,\zb)}{|\zb-\xb|^{d-2}}\,.
\end{equation}
This reduces to Eq.~16 of~\cite{Lavelle:1998dv} for $d=3$. The gauge
invariant, equal-time, free propagator in~(\ref{vd}), in $d+1$ dimensions, is given by
\begin{equation}\label{propd}
\bra{0}A^{\mathrm{T}}_j(\zb)A^{\mathrm{T}}_k(\wb)\ket{0}
=\frac{\Gamma(\frac{d+1}2)}{2\pi^{\frac{d+1}2}}
\frac{(z-w)_j(z-w)_k}{|\zb-\wb|^{d+1}}\,,
\end{equation}
which can be understood as the Coulomb propagator for the spatial
components or, in the Lorentz class of gauges, as the $\xi=-1$
propagator (where $\xi=1$ corresponds to Feynman gauge). This
last identification, based upon the requirement of the propagator being
transverse to $\partial_i^x$,
allows us to use the computational power of covariant gauges. The
dimension-independence of this gauge should be contrasted with the
Yennie gauge, where the propagator is transverse to
momentum derivatives \emph{only} in $d=3$.

Combining these expressions results in a standard, finite integral which may
be straightforwardly evaluated in $d$ dimensions. In $2+1$
dimensions, after Fourier transforming, we have
\begin{equation}
 V^{(4)}_\mathrm{anti}(q)=-{3\over2}g^4 C_{A}C_{F}
\frac1{\qb^2}
\inttwo{l}{1\over|\lb|(\lb-\qb)^2}\left(
1-{(\qb\cdot\lb)^2\over \qb^2\,\lb^2}
\right)\,.
\end{equation}
This is, of course, a \emph{finite} integral. From this we can rapidly show
that the anti-screening contribution to the potential at order
$g^4$ is given by
\begin{equation}\label{antis}
V_\mathrm{anti}^{(4)}(q)=- g^4 C_{F}
C_{A}{3\over4\pi}{1\over|\qb|^3}\,.
\end{equation}
Comparing this with the full potential in $2+1$ dimensions at this
order~(\ref{vtwolow}) we see first of all that the factors of
$\pi$ do not agree.
However,
the energy is still higher than the total result of
(\ref{vtwolow}) so it is still a physically acceptable result. We
now want to show how the screening contribution supplies the
different $\pi$ factors needed to lower the energy to the final
physical result. We
will, therefore,  now independently calculate the screening contribution.

We shall follow an approach to the potential which has been
presented by Gribov~\cite{Gribov:1977mi} and by Drell~\cite{Drell:1981gu}. Working now in
Coulomb gauge, the Hamiltonian is
\begin{equation}\label{Ham}
H={1\over2} \int
\mathrm{d}^2x\left(\left(\Eb^a_{\mathrm{T}}\right)^2+\left(\Bb^a
\right)^2-\phi^{a}\nabla^2\phi^{a}\right)\,,
\end{equation}
where we have decomposed the chromoelectric field into transverse
and longitudinal components, $\Eb^a=\Eb_T^a-\nablab\phi^a$ and
summation over colour is understood.
Gauss' law tells us that $\phi$ is related to the static matter
sources, $\rho$, and the gluonic fields by
\begin{equation}
\nabla^2\phi^a=g\left(\rho^a-f_{abc}\Ab^b\cdot\Eb^c\right),
\end{equation}
from which we can obtain the following equation up to order $g^3$,
which is all we shall require in this letter:
\begin{equation}\label{gauss}
\nabla^2\phi^a=\left\{
g\de^{ae}+g^2f_{abe}{\Ab^b\cdot\nablab}\frac1{\nabla^{2}}+g^3f_{abc}f_{cde}{\Ab^b
\cdot
\nablab}\frac1{\nabla^{2}}{\Ab^d\cdot\nablab}\frac1{\nabla^{2}}
\right\}\left(\rho^e-f_{egh}\Ab^g\cdot\Eb_\mathrm{T}^h\right) \,,
\end{equation}
where $\nabla^2$ acts on whatever is on its right.
We take the  sources for simplicity to have the form
$\rho^a=\rho_{q}^a+\rho_{\bar q}^a$ where
$\rho_{q}^a(\xb)=t^a_{q}\de^3(\xb)$, $\rho_{\bar q}^a(\xb)=
t^a_{\bar q}\de^3(\xb-\rb)$. Here we assume that $t^a_{q}$ and $t^a_{\bar q}$ are the
colour charges of a heavy (classical) quark $q_{i}$ and antiquark
$\bar q_{j}$ in a normalized colour singlet state
$|\Psi\rangle=N^{-1/2}|q_{i}\rangle|\bar q_{i}\rangle$. Hence the
colour factor becomes
\begin{equation}
 t^a_{q}t^a_{\bar q}=-{1\over N}\langle q_{i}|Q^a|q_{j}\rangle
 \langle \bar q_{i}|Q^a|\bar q_{j}\rangle={1\over N}\tr(T^a T^a)
 =-C_{F},
\end{equation}
where $Q^a$ is the colour charge operator and the anti-Hermitian generators
$T^a$ are in the fundamental representation of $SU(N)$. The heavy
quark and antiquark are again separated by $\rb$. The sources only
enter the Hamiltonian~(\ref{Ham}) in the last term, so the $\rb$
dependent term here  gives the potential between them.

Let us first explain how the lowest order result may be recovered in this approach.
The relevant term in the Hamiltonian is
\begin{equation}
-{1\over2}\int \mathrm{d}^2x\, \phi^a\nabla^2\phi^a=-g^2\int \mathrm{d}^2x\,
\rho_{\bar q}(x)\frac1{\nabla^2}\rho_{q}(x)\,,
\end{equation}
where, as we are only interested in the potential, we have dropped
separation independent terms.  We may now evaluate the
 expectation value between the gluonic vacuum
states. Expressing the delta functions as
Fourier transforms and trivially performing the spatial integral,
we obtain
\begin{equation}
V(q)=- g^2 C_{F} \frac1{\qb^2},
\end{equation}
from which we can read off the three dimensional generalisation of
the Coulomb interaction, i.e., the lowest order term in
(\ref{vtwolow}).

We may now proceed to the $g^4$ contributions. From time independent
perturbation theory we may write it as the sum of anti-screening
and screening effects, $V^{(4)}(r)=
V^{(4)}_\mathrm{anti}(r)+ V^{(4)}_\mathrm{scr}(r)$, where
\begin{eqnarray}
 V^{(4)}_\mathrm{anti}(r)&=&-{1\over2}\int \mathrm{d}^2 x\,\langle
0|\phi^a\nabla^2\phi^a|0\rangle,\label{eb27.1.00a}\\
V^{(4)}_\mathrm{scr}(r)&=&{-1\over4}\sum_{n\not=0}{1\over E_{n}}\int \mathrm{d}^2
x\,\langle               \label{screen}
0|\phi^a\nabla^2\phi^a|n\rangle
\int \mathrm{d}^2 x\,\langle
n|\phi^b\nabla^2\phi^b|0\rangle
\,,
\end{eqnarray}
and $E_{n}$ is the energy of the state $|n\rangle$.
In the second term it is sufficient, at this order, to sum over a
complete set of intermediate states of two transverse gluons.

These two terms again
represent the two distinct physical interactions that occur in QCD: the
first is the non-abelian generalisation of the Coulombic
interaction, while the second describes the exchange of physical,
transverse gluons. This second term, since it comes from the
exchange of physical quanta, represents the expected screening
part of the potential which lowers the interaction energy.

Let us first verify that this anti-screening contribution agrees
with our previous result. At this
order we need to retain terms up to order $g^3$ in $\phi$. From
(\ref{gauss}) we obtain three identical contributions
\begin{equation}
 V^{(4)}_\mathrm{anti}(r)=-{3\over2}g^4f_{abc}f_{cde}\int
\mathrm{d}^2x\,\langle0|\rho^a{1\over\nabla^2}\Ab^b
\cdot
\nablab\frac1{\nabla^{2}} \Ab^d\cdot\nablab\frac1{\nabla^{2}}
\rho^e|0\rangle \,.
\end{equation}
This result is precisely Eq.~\ref{vd} as expected.

To find the screening contribution, we
now need to insert physical two gluon states into (\ref{screen}).
The sum over
such states then becomes a sum over colour ($e$, $f$) and
helicity ($\la$, $\si$) and an integral over momenta. Explicitly we
have:
\begin{equation}
\sum_{n={\rm 2~gluon}}|n\rangle\langle n |=
{1\over2}\sum_{e\,f}\sum_{\la\,\si}\int \mathrm{d}^2k\int \mathrm{d}^2l\;
a^\dagger_{e}(\la,\kb)a^\dagger_{f}(\si,\lb)|0\rangle\langle
0|a_{f}(\si,\lb)a_{e}(\la,\kb) \,.
\end{equation}
The only terms from (\ref{gauss})
that can contribute to these transverse states are given by
$\phi^a\nabla^2\phi^a\to   -2 g^2 f_{abc}\, \rho^a\nabla^{-2}\Ab^b
\cdot\Eb_\mathrm{T}^c$. We may thus write
\begin{equation}
 V^{(4)}_\mathrm{scr}(r)=-2g^4\kern-.5cm\sum_{n={\rm 2~gluon}}{1\over E_{n}}
\int \! \mathrm{d}^2 x f_{abc}\, \langle0|{\rho^a_{q}}\frac1{\nabla^{2}}\Ab^b
\cdot\Eb_\mathrm{T}^c|n\rangle
\int\! \mathrm{d}^2 w f_{def}\, \langle n|{\rho^d_{\bar q}}\frac1{\nabla^{2}}\Ab^e
\cdot\Eb_\mathrm{T}^f|0\rangle\,,
\end{equation}
and, in Coulomb gauge, we may identify the transverse electric
field with $-\dot A_i$.
 Using the standard commutator, $[a_b(\la,\kb),a_{c}^\dagger(\si,\lb)]
 =\de_{bc}\de_{\la\si}\de^3(\kb-\lb)$ we
then rapidly obtain
\begin{eqnarray}
 V^{(4)}_\mathrm{scr}(q)&=&{g^4\over4}C_{A}C_{F}{1\over|\qb|^4}
\inttwo{l}\int
\mathrm{d}^2k{\de^3(\qb-\kb-\lb)(|\lb|-|\kb|)^2\over|\lb|\,|\kb|\,
(|\lb|+|\kb|)}\nonumber\\
&&\qquad\qquad\times\sum_{\la}\ep^i(\la,\kb)\ep^j(\la,\kb)
\sum_{\si}\ep^i(\si,\lb)\ep^j(\si,\lb)\,,
\end{eqnarray}
where we have already carried out the $\xb$ integral and some trivial
momentum
integrals.
We now
exploit the Coulomb gauge relation
\begin{equation}
\sum_{\la}\ep^i(\la,\kb)\ep^j(\la,\kb)=\de^{ij}-{k^ik^j\over \kb^2}\,,
\end{equation}
to arrive at the final expression for the screening contribution
\begin{equation}
 V^{(4)}_\mathrm{scr}(q)={g^4\over4}C_{A}C_{F}\frac1{|\qb|^4}
J(q)\,,
\end{equation}
where
\begin{equation}
 J(q)=\inttwo{l}{(|\lb|-|\qb-\lb|)^2\over|\lb|\,|\qb-\lb|\,
 (|\lb|+|\qb-\lb|)}\left\{
 1-{\qb^2\over(\qb-\lb)^2}+{(\qb\cdot\lb)^2\over \lb^2\,(\qb-\lb)^2}
 \right\}\,,
\end{equation}
The term $|\lb|+|\qb-\lb|$ in the denominator is unusual and makes this
a
more difficult integral\footnote{We need to calculate the full,
finite integral, while in $3+1$ dimensions, where we have the
same integrand, we only need to extract the logarithmic
divergence, so this  denominator
term effectively reduces to $2|\lb|$ and the
calculation is trivial.}.

To evaluate this integral it is convenient to go into polar
co-ordinates, $\rho,\theta$, and then make the change of variables:
$\rho=(\tau^2-1)/[2(\tau-\cos\theta)]$. The angular integral may then
be performed and afterwards it is not too difficult to evaluate
the integral over $\tau$. The result is
\begin{equation}
 J(q)=\left(-{7\over8}+{3\over\pi}\right)|\qb|  \,.
\end{equation}
 We so obtain for the screening
contribution to the interquark potential in $2+1$ dimensions
\begin{equation}
 V^{(4)}_\mathrm{scr}(q)=-{g^4}C_{F}C_{A}
{1\over4|\qb|^3}
\left({7\over8}-{3\over\pi}\right)
\,.
\end{equation}
This, together with the anti-screening result~(\ref{antis}),
gives us the total order $g^4$ contribution to the interquark
potential in $2+1$ dimensions:
\begin{equation}
 V^{(4)}(q)=-{g^4}C_{F}C_{A}
{1\over4|\qb|^3}
\left[ {3\over\pi}-\left({3\over\pi}-{7\over8}\right)
\right]
\,.
\end{equation}
We see that, as expected, the sum of the dominant anti-screening
contribution and this screening term gives exactly the correct
result for the total potential~(\ref{vtwolow}). As had to be the
case, the various factors of $\pi$ have combined to give one
overall factor.  This physical
decomposition cannot be seen in the Wilson loop approach. There
only the contributions from different classes of diagrams can be
distinguished, however, they are gauge dependent (and all have the
same $\pi$ factors).

The relative numerical weighting of the screening and
anti-screening contributions to the potential is now $8.37\%$. This is remarkably
within one part in a hundred of
the split in $3+1$ dimensions!

\smallskip

There is a pressing need for a detailed understanding of the
structure of the forces in QCD. In this paper we have calculated
the screening and anti-screening contributions to the static
inter-quark potential in $2+1$ dimensions:
\begin{equation}
V(r)=\frac{g^2C_F}{2\pi}\ln (g^2 r)  +\frac{g^4C_FC_A}{8\pi}
\left[ {3\over\pi}-\left({3\over\pi}-{7\over8}\right)
\right]r
\,.
\end{equation}
This calculation is of interest in itself, given that the beta function now vanishes, in that it
supports the idea that the $2+1$ theory models many of the important features of full QCD.
We have seen that, to a first approximation, it
is safe to neglect gluonic screening effects  in $2+1$ dimensions.
Additionally we note that it is important to understand  the $2+1$ dimensional theory,
since it is related by dimensional reduction to the high temperature
limit of QCD (although this identification is not direct~\cite{Kajantie:1996dw}).
The physical decomposition that we have calculated
exhibits a curious mathematical property (the differing
transcendental factors)  and the unexpected physical behaviour that the relative weights of
screening and anti-screening are nearly identical in both $2+1$ and $3+1$ dimensions.

We note that the one loop anti-screening coefficient of the linearly rising term
is only 1.35 times the lattice result in 2$+$1
dimension~\cite{Teper:1998te}. It is not clear to us why there is
such good agreement, nor how a linear potential emerges from the
lattice when higher perturbative corrections will have the form of a
power series in $r$.

We now want to extend this work in various directions. The
 separation
into screening and anti-screening is not known at
higher orders in the coupling. As noted by Drell~\cite{Drell:1981gu},
the method used in the latter part of this paper does
not easily lend itself to such calculations. Our approach, based on a
manifestly
gauge invariant construction of quarks and gluons, can be readily extended
to higher orders (see the appendix of~\cite{Lavelle:1997ty}). Indeed we have
previously shown~\cite{Bagan:1998kg} in QED
that such dressed fields have infra-red finite on-shell Green's functions at all
orders in perturbation theory. (The same  decomposition of the dressing into a minimal
and a separately gauge
invariant part is also reflected in the infra-red structures of QED.)  We are
thus in the process of calculating the, hitherto unknown,
decomposition of the potential into screening and anti-screening
effects at order $g^6$ in both $3+1$ and $2+1$ dimensions.
 Another important
extension of this work is to repeat the $3+1$ calculation at
finite temperature. The results of this letter could be taken as
indicating that the anti-screening/screening decomposition is
insensitive to the temperature. If this is indeed the case, one
needs to discover what aspect of strong interaction physics
underlies this remarkable property.

\medskip

\no\textbf{Acknowledgements:} This work was supported by the
British Council/Spanish Education Ministry \textit{Acciones
Integradas} grant no.\  1801/HB1997-0141. We thank S.\ Drell,
Y.\ Schr\"oder and S.\ Tanimura for
correspondence, and Robin Horan and Ramon Mu\~noz-Tapia  for discussions.


\end{document}